# Semantics and Conversations for an Agent Communication Language [*]

Yannis Labrou and Tim Finin
Computer Science and Electrical Engineering Department,
University of Maryland, Baltimore County, Baltimore, MD 21250, USA

## Abstract

We address the issues of semantics and conversations for *agent communication languages* and the Knowledge Query Manipulation Language (KQML) in particular. Based on ideas from speech act theory, we present a semantic description for KQML that associates "cognitive" states of the agent with the use of the language's primitives (performatives). We have used this approach to describe the semantics for the whole set of *reserved* KQML performatives. Building on the semantics, we devise the conversation policies, *i.e.*, a formal description of how KQML performatives may be combined into KQML exchanges (conversations), using a *Definite Clause Grammar*. Our research offers methods for a speech act theory-based semantic description of a language of communication acts and for the specification of the protocols associated with these acts. Languages of communication acts address the issue of communication among software applications at a level of abstraction that is useful to the emerging *software agents* paradigm.

## 1 Introduction

Communication among software agents [Petrie, 1996; Nwana, 1996] is an essential property of agency [Wooldridge and Jennings, 1995]. Agent communication languages allow agents to effectively communicate and exchange *knowledge* with other agents despite differences in hardware platforms, operating systems, architectures, programming languages and representation and reasoning systems. We view an agent communication language as the medium through which the attitudes regarding the content of the exchange between agents are communicated; it suggests whether the content of the communication is an assertion, a request, a query, *etc.*

Knowledge Query and Manipulation Language (KQML) is such a language; it consists of primitives (called *performatives*) that express attitudes regarding the content of the exchange and allow agents to communicate such attitudes to other agents and find other agents suitable to process their requests. Our research provides semantics for KQML along with a framework for the semantic description of KQML-like languages for agent communication. We also address the issue of conversations, *i.e.*, of sequences of causally-related messages in exchanges between agents and present a method for the specification of conversations (*conversation policies*).

After an introduction to KQML, we describe our semantic framework and give the semantics for a small set of KQML performatives. We follow with our method for describing the protocols (conversations) associated with the primitives and present the resulting conversations for our set of performatives. [1] We end by summarizing our contributions regarding the semantics and the specification of the conversations.

## 2 KQML for Agent Communication

KQML is an abstraction, a collection of *communication* primitives (message types) that express an *attitude* regarding the actual expression being exchanged, along with the assumptions of a simple model for inter-agent communication and an abstract design for KQML-speaking agents. There is no such thing as an *implementation* of KQML, *per se*, meaning that KQML is not an *interpreted* or *compiled* language that is offered in some hardware platform or an abstract machine. Agents *speak* KQML in the sense that they use those primitives, this *library of communication acts*, with their reserved *meaning*. The application programmer is expected to provide code that processes each one of the performatives for the agent's language or knowledge representation framework. This is a KQML message:

```
(ask-if  :sender     A
         :receiver   B
         :language   prolog
         :ontology   bible-genealogy
         :reply-with id1
         :content    ``spouse(adam,eve)'' )
```

---

[*]This work was supported in part by the Air Force Office of Scientific Research under contract F49620–92–J–0174, and the Advanced Research Projects Agency monitored under USAF contracts F30602–93–C–0177 and F30602–93–C–0028 by Rome Laboratory.

[1]Specifications for the full set of KQML performatives and associated policies are available in [reference omitted].

In KQML terminology, *ask-if* is a *performative*.[2] The value of the :content is an expression in some language or another KQML message and represents the content of the communication act. The other parameters (*keywords*) introduce values that provide a context for the interpretation of the :content and hold information to facilitate the processing of the message. In this example, *A* is querying *B* (these are symbolic names for agents[3]), in *Prolog* (the :language), about the truth status of *spouse(adam,eve)*. Any response to this KQML message will be identified by *id1* (the :reply-with). The ontology [4] *bible-genealogy* may provide additional information for the interpretation of the :content. In an environment of KQML-speaking agents there are agents called *facilitators* (*mediators* or *brokers* [Decker et al., 1996]) denote similarly intended agents to whom agents *advertise* their services and ask for assistance in finding other agents that can provide services for them.

Our goal is to provide a semantic description for the language in a way that captures all the intuitions expressed in its existing documentation [ARPA Knowledge Sharing Initiative, 1993]. The lack of semantics for KQML has been a long-standing problem of KQML. Moreover, although agents engage into extended interactions with other agents (conversations), conversations is an issue that has received little attention with respect to KQML, or other agent communication languages (the few notable exceptions are [Barbuceanu and Fox, 1995; Kuwabara, 1995; Bradshaw et al., 1996; Parunak, 1996]). Building on the semantic description we explore the issue of specifying KQML conversations in a formal manner.

## 3 A Framework for the Semantics

We treat KQML performatives as speech acts. We adopt the descriptive framework for speech acts and particularly illocutionary acts suggested by Searle [Searle, 1969; Searle and Vanderveken, 1985]. The semantic approach we propose uses expressions, that suggest the minimum set of preconditions and postconditions that govern the use of a performative, along with conditions that suggest the final state for the *successful* performance of the performative; these expressions describe the states of the agents involved in an exchange and use propositional attitudes like *belief, knowledge, desire*, and *intention* (this *intentional description* of an agent is only intended as a way of viewing the agent) which have the following reserved meaning:

1. BEL, as in BEL(A,P), which has the meaning that *P* is (or can be proven) true for A. *P* is an expression in the native language of agent *A*.[5]
2. KNOW, as in KNOW(A,S), expresses knowledge for *S*, where *S* is a state description (the same holds for the following two operators).
3. WANT, as in WANT(A,S), to mean that agent A desires the cognitive state (or action) described by *S*, to occur in the future.
4. INT, as in INT(A,S), to mean that A has every intention of doing *S* and thus is committed to a course of action towards achieving *S* in the future.

We also introduce two instances of actions:

1. PROC(A,M) refers to the action of *A* processing the KQML message *M*. Every message after being *received* is *processed*, in the sense that it is a valid KQML message and the piece of code designated with processing the performative for the application indeed processes it. PROC(A,M) does not guarantee proper processing of the message (or conformance of the code with the semantic description).
2. SENDMSG(A,B,M) refers to the action of *A* sending the KQML message **M** to *B*.

For an agent *A* it is BEL(A,P) if and only if *P* is true (in the *model-theoretic* sense) for *A*; we do not assume any axioms for BEL. Roughly, KNOW, WANT and INT stand for the psychological states of knowledge, desire and intention, respectively. All three take an agent's state description (either a cognitive state or an action) as their arguments. An agent can KNOW an expression that refers to the agent's own state or some other agent's state description if it has been communicated to it. So, KNOW(A,BEL(B,"foo(a,b)")) is valid, if BEL(B,"foo(a,b)") has been communicated to *A* with some message, but KNOW(A,"foo(a,b)") is not valid because "foo(A,B)" is not a state description. Researchers have grappled for years with the problem of formally capturing the notions of *desire* and *intention*. Various formalizations exist but none is considered a definitive one. We do not adopt a particular one neither we offer a formalization of our own. It is our belief that any of the existing formalizations would accommodate the modest use of WANT and INT in our framework.

Our semantic description, which includes expressions with the mental attitudes and actions we described, provides the following: **(1)** a natural language description of the performative's intuitive meaning; **(2)** an expression which describes the content of the illocutionary act and serves as a formalization of the natural language description; **(3)** preconditions that indicate the necessary state for an agent in order to send a performative (**Pre(A)**) and for the receiver to accept it and successfully process it (**Pre(B)**); if the preconditions do not hold a *error* or *sorry* will be the most likely response; **(4)** postconditions that describe the states of both interlocutors after the *successful* utterance of a performative (by the sender) and after the receipt and processing (but before a counter utterance) of a message (by the receiver); the postconditions (**Post(A)** and **Post(B)**, respectively) hold unless a *sorry* or an *error* is sent as a *response* in order to suggest the unsuccessful processing of the message; **(5)** a completion condition for the performative (**Completion**) that indicates the final state, after possibly a conversation has taken place and the intention suggested by the performative that started the conversation, has been fulfilled; and **(6)** any explanatory comments that might be helpful. the performative.

---
[2] The term was first coined by Austin [Austin, 1962], to suggest that some verbs can be uttered so that they perform some action.

[3] We will use the term *agents* to indiscriminately refer to all kinds of KQML-speaking programs and applications.

[4] An ontology is a repository of semantic and primarily pragmatic knowledge over a certain domain.

[5] The *native* language of the application may or may not have modal operators but we do not assume any, here.

## 4 Semantics for KQML Performatives

We present the semantics for five KQML performatives (*advertise*, *ask-if*, *tell*, *sorry* and *broker-one*) which can support some interesting agent conversations and illustrate our approach. [6] We first introduce our notation. For a KQML message **performative(A,B,X)**, **A** is the `:sender`, **B** is the `:receiver` and **X** is the `:content` of the performative (KQML message). Occasionally we use **M** to refer to an instance of a KQML message. Capital-case letters from the beginning of the alphabet (*e.g.*, $A$, $B$, etc.) are agents' names and letters towards the end of the alphabet (*e.g.*, $X,Y,Z$) are propositional contents of performatives. All underscores (_) are unnamed, universally quantified variables (they stand for performative parameters that do not have values in the KQML message). Capital case letters preceded by a question mark (?), *e.g.*, ?$B$, are existentially quantified variables.

All expressions in our language denote agents' states. Agents' states are either actions that have occurred (PROC and SENDMSG) or agents' mental states (BEL, KNOW, WANT or INT). Conjunctions ($\wedge$) and disjunctions ($\vee$) of expressions that stand for agents' states are agent's states, also, but we do not allow $\wedge$ and $\vee$ in the scope of KNOW, WANT and INT. Propositions in the agent's native language can only appear in the scope of BEL and BEL can only take such a proposition as its argument. BEL, KNOW, WANT, INT and actions can be used as arguments for KNOW (actions should then be interpreted as actions that have already happened). WANT and INT can only use KNOW or an action as ar-guments. When actions are arguments of WANT or INT, they are actions to take place in the future.

---

[6]Semantics for the complete set appear in [Labrou, 1996].

A negation of a mental state is taken to mean that the mental state does not hold in the sense that it should not be inferred (we will use the symbol not). When ¬ qualifies BEL, *e.g.*, ¬ (BEL(A,X)), it is taken to mean that the `:content` expression $X$ is not true for agent $A$, *i.e.*, it is not provable in $A$'s knowledge base. Obviously, what "not provable" means is going to depend on the details of the particular agent system, for which we want to make no assumptions.

**advertise(A,B,M)**
1. A states to B that A can and will process the message $M$ from B, if it receives one (A commits itself to such a course of action).
2. INT(A,PROC(A,M))
   where $M$ is the KQML message **performative_name(B,A,X)**.
3. **Pre(A)**: INT(A,PROC(A,M))
   **Pre(B)**: NONE
4. **Post(A)**:
   KNOW(A,KNOW(B,INT(A,PROC(A,M))))
   **Post(B)**: KNOW(B,INT(A,PROC(A,M)))
5. **Completion**: KNOW(B,INT(A,PROC(A,M)))
6. An *advertise* is a commisive act, in the sense that it commits its sender to process $M$, as suggested by the announcement of the intention to process. If B is a *facilitator* then B is interchangeable (in the semantic description) with the name of any agent the facilitator knows about.

**ask-if(A,B,X)**
1. A wants to know what B believes regarding the truth status of the content $X$.
2. WANT(A,KNOW(A,S))
   where $S$ may be any of BEL(B,X), or ¬(BEL(B,X)).
3. **Pre(A)**: WANT(A,KNOW(A,S)) $\wedge$ KNOW(A,INT(B,PROC(B,M)))
   where $M$ is **ask-if(A,B,X)**
   **Pre(B)**: INT(B,PROC(B,M))
4. **Post(A)**: INT(A,KNOW(A,S))
   **Post(B)**: KNOW(B,WANT(A,KNOW(A,S)))
5. **Completion**: KNOW(A,$S'$) )
   where $S'$ is either BEL(B,X) or ¬(BEL(B,X)), but not necessarily the same instantiation of $S$ that appears in $Post(A)$, for example.
6. **Pre(A)** and **Pre(B)** suggest that a proper advertisement is needed to establish them (see *advertise* and our comments in Section 7).

**tell(A,B,X)**
1. A states to B that A believes the content to be true.
2. BEL(A,X)

3. **Pre(A)**: BEL(A,X) ∧ KNOW(A,WANT(B,KNOW-(B,S)))
   **Pre(B)**: INT(B,KNOW(B,S))
   where $S$ may be any of BEL(B,X), or ¬(BEL(B,X)).
4. **Post(A)**: KNOW(A,KNOW(B,BEL(A,X)))
   **Post(B)**: KNOW(B,BEL(A,X))
5. **Completion**: KNOW(B,BEL(A,X))
6. The completion condition holds, unless a *sorry* or *error* suggests B's inability to acknowledge the *tell* properly, as is the case with any other performative.

**sorry(A,B,Id)**

1. A states to B that although it processed the message, it has no (possibly further) response to provide to the KQML message $M$ identified by the :reply-with value **Id** (some message identifier).
2. PROC(A,M)
3. **Pre(A)**: PROC(A,M)
   **Pre(B)**: SENDMSG(B,A,M)
4. **Post(A)**: KNOW(A,KNOW(B,PROC(A,M))) ∧ not($Post_M(A)$),
   where $Post_M(A)$ is the **Post(A)** for message $M$.
   **Post(B)**: KNOW(B,PROC(A,M)) ∧ not($Post_M(B)$)
5. **Completion**: KNOW(B,PROC(A,M))
6. The postconditions and completion conditions do not hold, even though A dispatched the performative to the appropriate function, because A could not (or did not want) to come up with a response that would result to their satisfiability. The not should be taken to mean that the mental state it qualifies should not be inferred to be true as a *result* of this particular message. This does not mean that for example $Post_M(B)$ does not hold if it has already been established by a previous message; it is up to B to decide (perhaps after using additional information) if and how it wants to alter its internal state with respect to the *sorry*.

**broker-one(A,B,performative(A,_,X))**
Let $D$ be an agent such that CANPROC(D,performative(B,D,X))[7] and *performative* be a performative that entails a request (a *directive*); for the set of performatives presented here, only *ask-if* falls into this category. B sends **performative(B,D,X)** to $D$, receives some *response* (depending on the *performative*) from $D$, let us call it **response(D,B,X')**, and then $B$ sends to $A$ the message **forward(B,A,_,A,response(_,A,X'))**.[8]

Semantically this is a three-party situation. We break down the semantic description to the three (agent) pairs involved in the transaction.

**A and B** For $A$ and $B$, the semantics are **not** those of a **performative(A,B,X)**, meaning that $A$ is aware that whatever *response*, if any, comes from $B$ is merely an "echo" of the utterance of the broker-ed agent $D$. So, the semantics is:

1. A wants B (a broker) to send the :content of the *broker-one* to some agent that can process it and eventually forward the response of the broker-ed agent back to A.
2. WANT(A,SENDMSG(B,D,M))
   where $M$ is **performative(B,D,X)** and D is an agent such that CANPROC(D,M).
3. **Pre(A)**: WANT(A,SENDMSG(B,D,M))
   **Pre(B)**: B has to be a *facilitator*; an agent can be a facilitator if and only if it can process performatives like *broker-one*, although it is usually more helpful to ascribe facilitator status to an agent in advance, so that agents can know which agent to contact for such requests.
4. **Post(A)**: KNOW(A,SENDMSG(B,D,M))
   **Post(B)**: SENDMSG(B,D,M))
5. **Completion**: SENDMSG(B,A,forward(B,A,_,A,M'))
   where M' is the message **response(_,A,X')** generated by the broker-ed agent's response to B, *i.e.*, **response(D,B,X')**.
6. To offer an example, if the :content of the *broker-one* was **ask-if(A,_,X)**, A understands that the (possible) *response* **forward(B,A,_,A,tell(_,A,X))** does not imply that BEL$(B,X)$, since $D$'s response to B is wrapped in a *forward* and then sent to $A$. Also, $D$'s name is omitted in the *forward*, so A does not know $D$'s name.

**B and D** For $B$ and $D$ the semantics are those of **performative(B,D,X)**, meaning that as far as $D$ knows of, the exchange has the meaning and repercussions of **performative(B,D,X)** (and whatever additional responses) being exchanged between $B$ and $D$.

**A and D** For $A$ and $D$ the semantics are those of **performative(A,D,X)** (let us call it $M$) but with the major difference that this is an one-sided exchange. So, $Pre_M(D)$ and $Post_M(D)$ are empty because $D$ does not know that it has this exchange with $A$. Additionally, $A$ can have no prior knowledge (in $Pre_M(A)$) of its interlocutor's state. Finally, the applicable $Post_M(A)$ and $Completion_M$ lack the name of $D$. To show how this translates semantically, we present the semantics of **broker-one(A,B,ask-if(A,_,X))** for agent $A$ and the broker-ed agent $D$.

1. A wants to know what some other agent believes regarding the truth status of the content $X$.

---

[7]CANPROC, as in CANPROC(A,M), stands for "A being able to process message $M$." It is always the case that if **advertise(A,B,M)** then CANPROC(A,M), but it could very well be the case that CANPROC(A,M) may be inferred in other ways (this is to be provided or inferred by $B$). CANPROC is entirely different from PROC; CANPROC suggests ability to process and PROC suggest that the agent will process (or has already processed) a performative, in the sense that it will (or did) dispatch the message to the appropriate piece of code for handling.

[8]The performative *forward* is not presented here. Its meaning is basically the intuitive one and the four parameters :from, :to, :sender and :receiver refer respectively to the originator of the performative in the :content, the final destination, the :sender of the *forward* and the :receiver of the *forward*.

2. WANT(A,KNOW(A,S))
   where $S$ may be any of BEL(?D,X), or ¬(BEL(?D,X)).
3. **Pre(A)**: WANT(A,KNOW(A,S))
   **Pre(D)**: NONE
4. **Post(A)**: INT(A,KNOW(A,S))
   **Post(D)**: NONE
5. **Completion**: KNOW(A,$S'$) )
   where $S'$ is either BEL(?D,X) or ¬(BEL(?D,X)), but not necessarily the same instantiation of $S$ that appears in $Post(A)$, for example.
6. In effect, $D$'s identity remains unknown to $A$ and $D$ is unaware that $A$ knows its belief regarding the truth status of X.

## 5 Describing Conversations

A *conversation* is a sequence of KQML messages that belong to the same thread of interaction between two or possibly more agents. We assume some sort of (intuitive) causal relation between messages that are taken to belong in the same *conversation* and we use the :in-reply-to value as the indicator of such linkage. *Conversation policies* are rules that describe permissible *conversations* among KQML-speaking agents. The *conversation policies* that we provide do not describe *all* possible *conversations* because more complex interactions (and thus conversations) are possible between KQML-speaking agents. The conversations we present can be used as building blocks for more complex interactions.

We use the Definite Clause Grammars (DCGs) formalism for the specification of the *conversation policies* for the KQML performatives. DCGs extend Context Free Grammars (CFGs) in the following way [Perreira and Warren, 1986]: 1) **Non-terminals** may be compound terms (instead of just atoms as in the CFG case), and 2) the body of a rule may contain **procedural attachments**, written within "{" and "}" (in addition to terminals and non-terminals), that express extra conditions that must be satisfied for the rule to be valid. For example, a DCG rule might look like

**noun(N) —→ [W], {RootForm(W,N), is_noun(N)}**

with the possible meaning that "a phrase identified as the noun **N** may consist of the single word **W** ([**W**] is a terminal), where **N** is the root form of **W** and **N** is a noun" [Perreira and Warren, 1986].

### 5.1 DCGs & KQML conversation policies

*Conversation policies* describe both the sequences of KQML performatives and the constraints and dependencies on the values of the *reserved parameters* of the performatives involved in the conversations. In other words, we are not only interested in asserting that an *ask-if* might be followed by a *tell* (among other performatives) but we want to also capture constraints such as, the content$_{ask-if}$ being the same with the content$_{tell}$ or the reply−with$_{ask-if}$ being also the

in−reply−to$_{tell}$. The DCG we provide in the next section fully describes the above in a declarative fashion.

Each KQML message is a *terminal* in the DCG. A terminal is a list of the following values: performative_name, :sender, :receiver, :in-reply-to, :reply-with, :language, :ontology, IO (if IO is set to 1 the message is an incoming message and if it is set to 0 the current message is an outgoing message), :content, and whenever the :content is a performative itself, then the :content is going to be a list itself. *Terminals* are enclosed in "[" and "]", so a terminal in our DCG will look like: [[**ask-if,A,B,id1,id2,prolog,bar,foo(X,Y)**]] In the DCG we present here, we omit the :language and :ontology values (we take them to remain unchanged throughout the same conversation).

The conversation policies we present are tied to the semantics in the sense that changes in the semantic description would result to different conversation policies. Our conversation policies technically are not inferred from the semantic description, but they define the *minimal* set of conversations that are consistent with the semantics when following these heuristics:

- If a performative has *preconditions* for the sender, then it cannot start a conversation if these preconditions have to be established by a communication act (see *tell*).
- If the *completion condition(s)* for a performative are not not a subset of the postconditions, then a performative cannot end a conversation since further (communicative) action has to take place to establish the *completion condition(s)* (see *ask-if*).
- A performative may be preceded by a performative that can (partially) establish its preconditions (*e.g.*, a *tell* may be preceded by an *ask-if*; compare **Post(A)** for *ask-if* and **Pre(A)** for *tell*).

## 6 Converation Policies, in detail

We present a complete DCG for the set of performatives presented in Section 4. This is a subset of the full DCG that describes the whole set of conversation policies (see [Labrou, 1996]) and is intended as a demonstration of how our method may be used.

**ask-if, tell**

S →
    s(CC,P,S,R,IR,Rw,IO,C),
    {member(P,[advertise,broker-one])}
s(CC,ask-if,S,R,IR,Rw,IO,C) →
    [[ask-if ,S,R,IR,Rw,IO,C]] |
    [[ask-if ,S,R,IR,Rw,IO,C]], {OI is abs(1-IO)},
        r(CC,ask-if,S,R,_,Rw,OI,C)
r(CC,ask-if,R,S,_,IR,IO,C) →
    [[tell ,S,R,IR,Rw,IO,C]] |
    problem(CC,R,S,IR,_,IO)
The rules are organized into groups that describe the sub-dialogues that may start with a performative, or a group of them and are written so that any sequence of messages that is reachable from the start is also a conversation that will be accepted by the DCG. Note that there is no notion of a *complete* KQML conversation, although it might be possible to define such conversations in some cases. Rules might be called by other rules.

As a result, an *advertise* of an *ask-if* is a conversation; if a proper *ask-if* follows the *advertise*, the sequence of *advertise* and *ask-if* is a conversation; and finally, if an appropriate *tell* follows the *ask-if*, the resulting sequence of the three messages will be a conversation that the DCG will accept. The values of the various terminals and non-terminals define what an *appropriate* follow-up is, at any point of a KQML exchange. We use the following variables for the various tokens that appear in the DCG (symbols that start with a capital-case letter are variables and those that start with small-case letters are constants): CC stands for the *current conversation* that the DCG handles; P is the *performative_name*; S is the :sender; R is the :receiver; IR is the :in-reply-to value; Rw is the :reply-with; IO and OI are the variables that indicates if a message is an incoming or outgoing one (they only take the values 0 and 1 and always have complimentary values) ; C is the :content; and [] is the *empty string*.

We take the position that all starting points for conversations are *advertise* performatives and the *broker-one* performative (when sent to, or processed by facilitators). *Ask-if* may follow an *advertise* and may be responded to (in this KQML subset) with a *tell*.[9] The :in-reply-to value of the response must equal the :reply-with of the *ask-if* for all performatives that act as a *response* or a *follow-up* to some other performative. Also, notice that the :content of a response is the same as the :content of the querying performative in the case of the *ask-if*.

### sorry

problem(CC,R,S,IR,Rw,IO) →
    [[sorry ,S,R,IR,Rw,IO,[]]]

A *problematic* or a *non-positive* response, *i.e.*, a *sorry* (or an *error*, not included here) is always a possibility and those two performatives may follow almost any performative (except for another *sorry* or *error*).

### advertise

s(CC,advertise,S,R,_,Rw,IO,_) →
    { OI is abs(1-IO) },
    [[advertise,S,R,_,Rw,IO,[P1,R,S,Rw,_,OI,C1]]] ,
        {member(P1,[ask-if])},
        c_adv(CC,P1,S,R,Rw,_,OI,C1)
c_adv(CC,P,R,S,Rw_adv,_,IO,C) →
    s(CC,P,S,R,Rw_adv,_,IO,C) |

---
[9]A response with a *sorry* or *error* (not included in this set) is always a possibility of course.

    problem(CC,S,R,Rw_adv,_,IO) | []

The procedural attachment restricts the performatives that might appear in the :content of an *advertise*. The :content has the form of the expected follow-up to the *advertise*. This follow-up is given by the part of the DCG that starts the sub-dialogue for the embedded performative. Note that it is possible to have a *sorry* response to the *advertise* itself, as well to the follow-ups to the *advertise*.

### broker-one

s(CC,broker-one,S,R,IR,Rw,IO,C) →
    {OI is abs(1-IO)},
    [[broker-one,S,R,IR,Rw,IO,[P1,R,_R,Rw,Rw1,_,C1]]] ,
        {member(P1,[ask-if])},
        c_brk_one(CC,P1,S,R,Rw,Rw1,OI,C1)
c_brk_one(CC,P,R,S,Rw_brk,Rw,1,C) → [] |
    problem(CC,S,R,Rw_brk,_,1) |
    r(CC,P,Brk,R,_,Rw,1,C)
c_brk_one(CC,P,R,S,Rw_brk,Rw,0,C) → [] |
    problem(CC,S,R,Rw_brk,Rw,0) |
    s(CC,P,S,Brk,Rw_brk,Rw,0,C),
        c_brk_one1(CC,P,S,R,Brk,Rw_brk,Rw,0,C)
c_brk_one1(CC,P,S,R,Brk,Rw_brk,Rw,IO,C) → [] |
    {OI is abs(1-IO), last(CC,[P1,Brk,S,Rw,Rw1,OI,C1]),
      assert(send_MSG([forward,S,R,Rw_brk,Rw2,OI,
            [P1,_Brk,R,Rw_brk,Rw1,OI,C1]]))} |
    [[forward,S,R,Rw_brk,Rw2,OI,
        [P1,_Brk,R,Rw_brk,Rw1,OI,C1]]]

The *broker-one* performative presents an interesting case because it involves a three-party interaction. The receiver$_{broker-one}$ sends the content$_{broker-one}$ (with the appropriate values) to some other agent and then passes the response(s) to it to the sender$_{broker-one}$. The last part of this exchange can be done automatically with a procedural attachment in the DCG instead of being taken care of by the handler function for *broker-one*. As the c_brk_one1 rule suggests, a sub-dialogue (a new conversation) with the third agent starts and the response (or follow-up), *i.e.*, the last message in the conversation being handled by the DCG with the expected values for :sender and :in-reply-to, is sent to sender$_{broker-one}$ (this is the meaning of the procedural attachment in the c_brk_one1 rule, that makes reference to predicates that are not a part of the DCG).

If the **local** agent sent a *broker-one*, the message expected is the prescribed response or follow-up to the *performative* in the :content. Technically this message (or messages) will arrive wrapped in a *forward* but from the DCG point of view will be stripped from their "forwarding" packaging. This performative is a prime example of how complicated interactions might be composed from the simpler building blocks.

# 7 Discussion

The issue of semantics for communication acts has received a fair share of attention. Cohen and Lesveque suggest a model for rational agents [Cohen and Levesque, 1990], which uses a *possible-worlds* formalism, that can in turn be used as a framework for the semantic description of illocutionary acts [Cohen and Levesque, 1995; Smith and Cohen, 1996]. Sadek [Sadek, 1992] has also taken on a similar task of defining rational agency and defining communicative acts on top of it. Finally, Singh proposes a model of agency [Singh, 1993a], which differs from that of Cohen and uses it as a framework for the semantic treatment of speech acts [Singh, 1993b].

In contrast, we draw directly from a high-level speech act account, although the resulting preconditions-/postconditions framework is reminiscent of planning (but it could also be thought as operational semantics, *i.e.*, transitions on agents' states). Also, we provide no formal semantics (in a *possible-worlds* formalism or some similar framework) for the modal operators but we restrict the scope and use of these operators, so that they can be subsumed by similar modalities whose semantics could be provided by an intentional theory of agency. Apart from the complexity of possible-worlds–like formalisms which can be prohibiting for the intended audience of our semantic description that includes application developers that want to support KQML in their software agents, we want to avoid a tight coupling with a particular theory of agency. Another common element of the mentioned approaches is the strictly declarative definitions of the primitives. Instead, our preconditions, postconditions and completion conditions framework suggests a more operational approach which we hope will be useful to implementors that have to provide the code that processes the communication primitives.

By attempting a semantics for communication acts without a theory of agency, *i.e.*, formal semantics for the propositional attitudes (operators), we certainly give up interesting inferencing. For example, if an agent sends **tell(A,B,X)** and later **tell(A,B,X $\to$ Y)**, $B$ will not be able to infer that BEL(A,Y) (since we do not even assume a universal *weak S4* model for BEL) based on the KQML semantics alone. Nothing is lost though, because the additional information of the agent theory that holds for the agent can be supplied as part of the KQML exchange (*e.g.*, in the :ontology value of a KQML message) and subsequently taken into consideration for further inferencing. In the end, we trade a formal semantics for the propositional attitudes, which inevitably define a *model of agency* that is unlikely to be universal for all agents, for a simpler formalism and agent theory independence.

Objections may be raised regarding some of our choices regarding the meaning we chose to attribute to some of the performatives. Our semantics for *tell*, for example, suggest that an agent can not offer unsolicited information to some other agent. This can be easily amended by introducing another performative, let us call it *proactive-tell* which has the same semantic description as *tell* with the following difference: **Pre(A)** is BEL(A,X), and **Pre(B)** is empty. Similarly, an agent $A$ can send an *ask-if* to agent $B$ if and only if $A$ knows that $B$ is going to process such a request. Implicit in this choice, is our preference for a model where agents advertise their services so that other agents (with the help of *mediators* or *facilitators*) can find agents that can process requests for them. A "relaxed" version of *ask-if* can be introduced to allow for direct querying. The semantic description of this *proactive-ask-if* differs from that of *ask-if* as follows: **Pre(A)** is WANT(A,KNOW(A,S)), and **Pre(B)** is empty. Following KQML's tradition of an open standard, the KQML users' community should decide the performative names to be associated with whatever semantic description. Additionally, these two "new" performative could be starting points for conversations in our conversation policies.

Our description and implementation of the conversation policies using a DCG allows as to provide a description that would not be possible had we chosen a CFG or a Finite State Machine for the task. Another formalism that would probably provide us with the same flexibility is that of Augmented Transition Networks[10] (ATNs), but DCGs have the advantage that they can be expressed directly in a general purpose programming language like Prolog (in fact our DCG is a Prolog program). The conversation policies do not prescribe the only possible behavior for an agent but they rather define one which is consistent with the semantics. Such a specification is in no way a prescriptive one and thus does not constrain elaborate agents but it could be useful for simpler ones.

# 8 Conclusions

We have presented excerpts of a complete semantic description for the primitives in the agent communication language KQML. This specification uses a framework for the semantic description of KQML-like languages [11] for the linguistic communication among software agents along with a method for specifying the conversations that builds on our semantic description. We have used our approach to provide the semantics and conversation policies for the full set of KQML primitives and we have presented the framework and the semantic description along with the method and the conversation policies'

---

[10] Perreira and Warren claim that DCGs are at least as powerful of a formalism as ATNs ([Bates, 1979]), with DCGs having some considerable advantages over ATNs ([Perreira and Warren, 1986]).

[11] That is, languages of attitude-expressing communication primitives, modeled after speech acts.

specification for a handful of performatives.

The conversation policies present us with some attractive possibilities. They can be used to devise a software component that monitors an agent's incoming and outgoing messages and ensures that it only engages in valid KQML conversations of well-formed KQML messages. Such a component can keep track of an agent's multiple interactions (conversations) with other agents and offer ways to recover from unforeseen situations. Alternatively, one may view an agent as a collection of conversations that "unfold" concurrently as the agent interacts with other agents. Finally, the conversation policies can be used as building blocks for more complex interactions. In the end, we should keep in mind that agents do not use the primitives of a communication language statically, but in order to carry, often complex, interactions which the conversation policies can help describe.